\documentclass[12pt]{iopart}          
\usepackage{url}

\RequirePackage[T1]{fontenc}

\RequirePackage{multirow}
\RequirePackage{graphicx}
\RequirePackage{flushend}
\RequirePackage[colorlinks,citecolor=blue,urlcolor=blue,linkcolor=blue]{hyperref}

\usepackage{graphicx}  
\usepackage{dcolumn}   
\usepackage{bm}        
\usepackage{amssymb}   
\usepackage{feynmf}    
\usepackage{hyperref}  %
\usepackage{slashed}
\usepackage{color}
\usepackage{array}
\usepackage{caption}
\usepackage{lineno}
\usepackage{subcaption}
\expandafter\let\csname equation*\endcsname\relax
\expandafter\let\csname endequation*\endcsname\relax
\usepackage{amsmath}
\usepackage{booktabs}
\usepackage{multirow}
\usepackage{subcaption}
\usepackage{wrapfig}
\usepackage{dblfloatfix}
\usepackage{url}
\usepackage{relsize}
\usepackage[shortlabels]{enumitem}
\usepackage{parskip}
\usepackage[style=numeric-comp,backend=biber,sorting=none,
maxbibnames=99, minbibnames=99]{biblatex}
\newsavebox\mybox

\begin{document}

\title[TIGER]{TIGER: A Topology-Agnostic, Hierarchical Graph Network for Event Reconstruction}

\author{Nathalie Soybelman$^1$, Francesco A. Di Bello$^2$, Nilotpal Kakati$^1$, Eilam Gross$^1$}
\address{$^1$ Weizmann Institute of Science, Israel}
\address{$^2$ University of Genova and INFN Sezione di Genova, Italy\footnote{Now at the University of Pisa}}

\ead{nathalie.soybelman@weizmann.ac.il}
\vspace{10pt}
\begin{indented}
\item[]September 2025
\end{indented}

\begin{abstract}
Event reconstruction at the LHC, the task of assigning observed physics objects to their true origins, is a central challenge for precision measurements and searches.
Many existing machine learning approaches address this problem but rely on a single event topology, restricting their applicability to realistic analyses where multiple signal and background processes with different structures are present.
To overcome this, we present TIGER, a novel hierarchical graph network that is fundamentally topology-agnostic. By incorporating only the common underlying structure of sequential two-body decays, our model can reconstruct complex events without process-specific assumptions. This flexible architecture supports multi-task learning, enabling simultaneous event reconstruction and classification. TIGER thus provides a powerful and generalizable tool for physics analysis at the LHC.

\end{abstract}
\vspace{1pc }
\vspace{1pc}

\section{Introduction}
At the Large Hadron Collider (LHC)~\cite{lhc}, the reconstruction of collision events is a highly important and complex task. Since unstable particles decay in cascades before reaching the detector, only their resulting decay products are captured. The raw detector signals are first reconstructed into stable final-state particles by a subsequent chain of reconstruction algorithms, which are then clustered into high-level objects like jets~\cite{hgpflow1,hgpflow2,MLPF}. Our work addresses the final combinatorial challenge: accurately assigning these objects to their original "mother" particles. This is fundamental to nearly all particle physics analyses. However, in events involving multiple unstable particles, the large combinatorics become intractable for traditional methods such as $\chi^2$ and likelihood approaches~\cite{topograph, spanet0, spanet1,tt_atlas,tt_atlas_new,klfitter}.

To overcome these limitations, various machine learning approaches emerged in recent years, designed to tackle the combinatorial problem by embedding physical knowledge directly into their network architectures. Notable examples include Topograph~\cite{topograph}, a graph neural network that exploits the hierarchical decay structure of top quarks ($t\rightarrow bW\rightarrow bqq$) to reconstruct the $W$-boson before the top quark. Similarly, HyPER~\cite{hyper}, another graph neural network, utilises both edges and hyperedges to infer two- and three-particle decay assignments. Furthermore, SPANet~\cite{spanet0,spanet1,spanet2,spanet3} employs a transformer-based architecture with symmetry-preserving attention mechanisms to account for inherent symmetries among decay products. 

While the integration of physics knowledge is crucial for enhancing model performance and interpretability, it can also introduce significant limitations regarding generalizability. For instance, while HyPER and Topograph show strong performance in reconstructing $t\bar{t}$ events, their architectures will require modifications to address other decay topologies. In an realistic analysis, background events would be forced into the same predefined topology as the signal, limiting the performance of such specialised models.
Although the SPANet architecture offers greater generality and has been evaluated across various decay processes, it still necessitates explicit input of the event topology.

To address these shortcomings, we introduce a novel topology-agnostic approach that incorporates fundamental physical constraints without sacrificing generality or introducing bias. Our method is built on the observation that, in most cases, particles decay in at most two stages, each producing two daughter particles (e.g. $t \rightarrow Wb \rightarrow qq'b$, $H \rightarrow ZZ \rightarrow 4l$). These assumptions hold for nearly all processes studied at the LHC. However, certain cases such as trilinear Higgs decays, effective theories, including Fermi theory for low energy hadron decays, or possible new physics scenarios can involve three or more-prong decays or additional decay stages. In such situations, our method can be naturally extended, though exploring these generalizations lies beyond the scope of this work.

Importantly, our approach requires no prior knowledge of the event topology or the particles it comprises, making this a highly flexible method capable of simultaneously reconstructing and classifying both signal and background events across diverse event structures. In this work, we first demonstrate the model's reconstruction performance on fully hadronic $t\bar{t}$ and semileptonic $t\bar{t}H$ events. Subsequently, we evaluate its effectiveness in signal-versus-background discrimination by classifying between $t\bar{t}H$ and $t\bar{t} + bb$ events.

\section{Datasets}
This study utilises two distinct datasets, taken from the studies presented in~\cite{hyper} and~\cite{spanet2}. These datasets describe both fully hadronic and semileptonic event topologies.

The first dataset~\cite{hyper_dataset} comprises fully hadronic $t\bar{t}$ events. These events were generated with the following simulation chain: \textsc{MadGraph5\_amc@nlo} (v2.9.16) for matrix element generation, \textsc{Pythia8} (v8.306) for parton showering and hadronisation, and \textsc{Delphes} (v3.5.0) for detector simulation, employing the ATLAS detector card.

For jet reconstruction, the anti-$k_t$~\cite{anti-kt} algorithm was applied with a radius parameter of $R = 0.4$. A $p_T$-dependent tagging efficiency was used for $b$-tagging. Reconstructed jets were required to satisfy $p_T > 25$ GeV and $|\eta| < 2.4$. Events were selected if they contained a minimum of six jets, with at least two of these being $b$-tagged. The dataset is partitioned into approximately 7.9 million events for training, and 440,000 events each for validation and testing. Further details regarding this simulation can be found in~\cite{hyper}.

The second dataset\footnote{\url{https://mlphysics.ics.uci.edu/data/2023_spanet/}} focuses on semileptonic $t\bar{t}H$ events, where one top quark decays hadronically and the other semi-leptonically, and the Higgs boson decays into two $b$-jets. The simulation chain for these events involved \textsc{MadGraph5\_amc@nlo} (v3.2.0), \textsc{Pythia8} (v8.2), and \textsc{Delphes} (v3.4.2), configured with the CMS detector card.

Jets were clustered using the anti-$k_t$ algorithm with a radius parameter of $R=0.5$. Selected jets were required to have $p_T > 25$ GeV and $|\eta| < 2.5$. The same $b$-tagging procedure used for the fully hadronic $t\bar{t}$ dataset was applied. Electrons and muons were selected with the same $p_T$ and $\eta$ cuts. No specific requirements were imposed on the missing transverse energy ($E_{T}^\text{miss}$). Event selection required the presence of one lepton and a minimum of six jets. For the training and validation sets, a loose requirement of at least two $b$-tagged jets is applied, whereas the testing set follows a stricter criterion of at least four $b$-tagged jets. This dataset is allocated with approximately 9.9 million events for training, 520,000 for validation, and 240,000 for testing.

For the signal and background classification study, this $t\bar{t}H$ dataset was combined with a $t\bar{t}+bb$ sample, which represents the most dominant background for the described process. The $t\bar{t}+bb$ dataset adds 5 million events for training, 270,000 for validation, and 150,000 for testing.

\section{TIGER}
We introduce a novel method, TIGER (\textit{Topology-Independent Graph-based Event Reconstruction}), which employs a hierarchical, physics-inspired graph learning approach. 

\subsection{Network Architecture}

\begin{figure}[t]
    \centering
    \includegraphics[width=\linewidth]{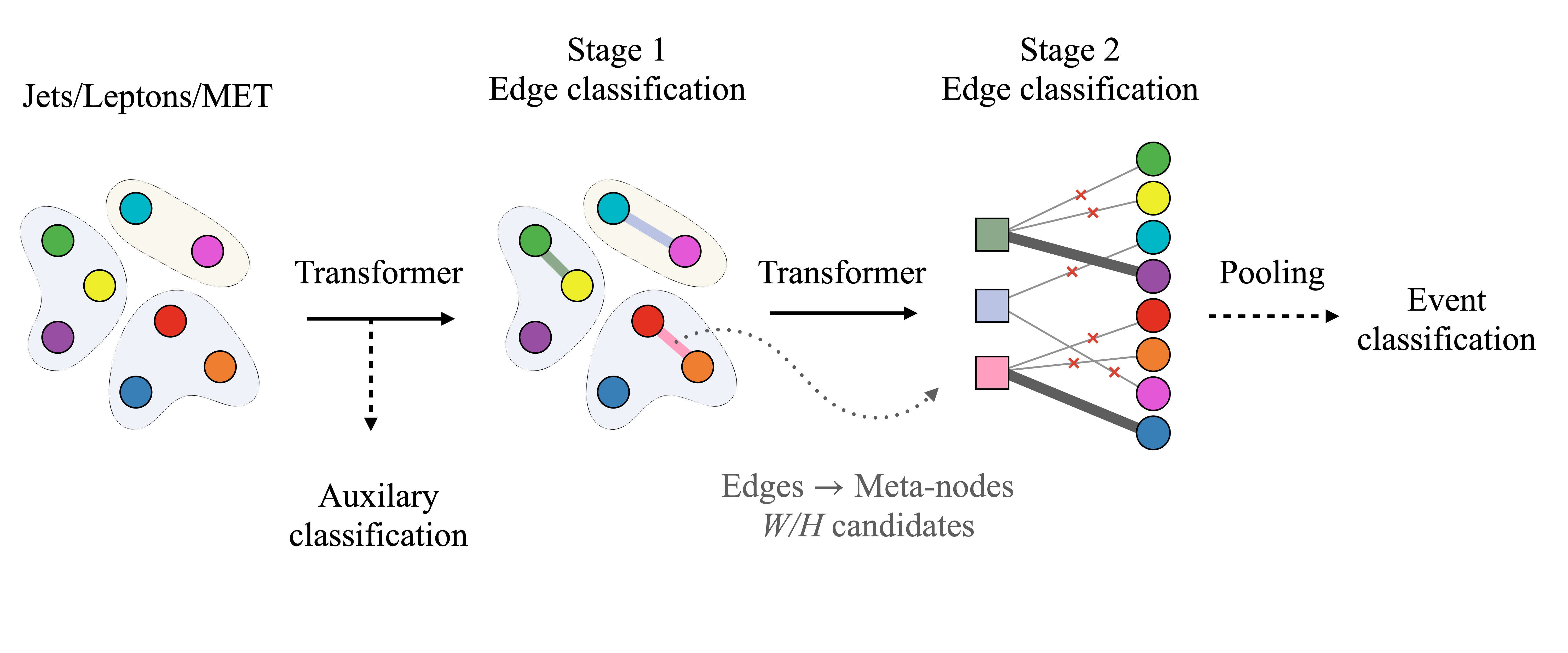}
    \caption{Overview of the TIGER architecture. After an initial transformer-based encoding, a two-stage hierarchical learning module reconstructs the event. The first stage identifies intermediate particles ($W,H$) as meta-nodes, while the second stage combines them with other nodes to form top quarks. An optional final block pools features from the graph for event classification.}
    \label{fig:architecture}
\end{figure}

The main building blocks of our model are graph neural networks and transformers, both have demonstrated broad applicability within particle physics, including tasks such as jet tagging~\cite{ftag_review, GN2,lorentz_equiv}, event classification~\cite{classification_fm}, anomaly detection~\cite{graph_anomaly}, reconstructions of particles~\cite{hgpflow1,hgpflow2, MLPF} or tracks~\cite{eggnet,egnn_tracking, hgnn_tracking,trackfitting, track_gnn}, as well as in fast simulations of detector responses~\cite{calograph}.

The overall network architecture consists of three primary components: encoding, hierarchical graph learning, and an optional event classification block. Each of these parts is described in detail below. An illustration of the complete architecture is provided in Figure~\ref{fig:architecture}.

\textbf{Encoding} The encoding stage begins by taking the input features of jets and leptons ($p_T$, $\eta$, $\phi$, $m$) and missing transverse energy (MET) ($E_T$, $\phi$). These features are then concatenated with their respective one-hot-encoded $b$-tag or lepton-tag information. The resulting feature vectors are then embedded into a hidden dimension using a separate Multi-Layer Perceptron (MLP) for each particle class. Following this embedding, all input objects are treated uniformly as nodes in the graph. This approach treats the MET as a particle, implicitly assuming it corresponds to a neutrino. While this simplification may not be entirely accurate from a physics perspective, as MET can arise from multiple neutrinos, new physics or detector issues, it preserves the model's simplicity and generality for the studies presented here. An alternative approach used in SPANet~\cite{spanet2}, where MET is treated as global event information, could be implemented, but was not found to be necessary for this work.

We then employ a diffusion transformer (DiT), an architecture originally proposed in~\cite{DiT}, to update these hidden representations. To provide contextual information, the mean of the hidden representations of all nodes, as well as the total number of objects in the event, are passed to the transformer. The updated hidden representations are subsequently passed to the hierarchical graph learning module.

Optionally, an auxiliary task can be performed at this stage to classify the origin of the particles. This task can be designed based on the expected signal and background processes, by considering all possibile particle origins and defining corresponding classification labels. For the specific case of the $t\bar{t}$ dataset, the classes are: not associated, associated with a reconstructable $W$, associated with a non-reconstructable $W$, $b$-jet from a reconstructable top, $b$-jet from a non-reconstructable top. For the semileptonic $t\bar{t}H$ dataset, these classes are naturally extended through the addition of the Higgs and the leptonic decay mode of the $W$. While the predictions from this auxiliary classification task are not directly used in the final reconstruction, we observe that it improves the overall network performance and interpretability.

\textbf{Hierarchical Graph Learning} This component of the architecture is loosely inspired by previous work~\cite{HG_pool, HG_diff_pool}, but has been substantially modified to incorporate physical principles. The hierarchical graph learning proceeds in two stages, corresponding to the two levels in the decay chain. 

In the first stage, a fully connected graph is constructed from the encoded input objects. Edge representations are formed by combining the node representations of each connected pair, which are then passed through an MLP for edge classification. The goal of this step is to determine whether two reconstructed objects originate from the same parent particle, and if so, to identify the type of that parent. Depending on the process under study, this can be formulated as either a binary or multi-class classification task. For example, in $t\bar{t}$ events, the task involves identifying whether two objects form a $W$ boson, while in $t\bar{t}H$ events, the classification extends to distinguishing between no connection, a $W$, or a Higgs boson.

In the second stage, the goal is to combine the intermediate particles identified in the first stage with other particles to form the final-state particles. To achieve this, a new graph is created with two types of nodes: the original nodes from the first graph and \textit{meta-nodes}, which represent the edges classified as the intermediate parent particles. Depending on the expected processes one can fully connect all initial nodes with the meta-nodes (e.g. for top-decays), the meta-nodes among themselves (e.g. diboson Higgs decays like $H\rightarrow WW$), or both. The following edge classification is conceptionally identical to the one in the first stage.

The meta-nodes are selected based on the reconstruction probability predicted in the first stage. A maximum of $k$ meta-nodes are chosen if their probability exceeds a certain threshold $p_\text{th}$, which are tunable hyperparameters. Notably, nodes that are incorporated into meta-nodes are not discarded during the training process. This is to prevent errors from the first stage from propagating and negatively impacting the performance of the second stage. For instance, if a $b$-jet is incorrectly included in a fake $W$, it would be unavailable for correct matching in the second stage. The model may also produce ambiguous assignments, such as two $W$ candidate predictions sharing a jet, or unphysical pairings where a meta-node is matched with one of its own constituents. To ensure training stability, these configurations are permitted to form but are discouraged by the loss. At the evaluation stage, a dedicated algorithm resolves these conflicts to guarantee a physically consistent output and prevent any double-counting of particles.

In cases where rare event topologies involving multiple decay stages or three-prong structures are expected, the approach can be generalized through an additional hierarchical level and a transition from edge to hyperedge representations.

\textbf{Event classification} The final, optional component of the architecture is an event classification block for signal versus background discrimination. The representation of the hierarchical graph is pooled by concatenating the mean of all nodes, the mean of all meta-nodes, the number of nodes, and the number of meta-nodes. This pooled representation is then passed through an MLP to perform a binary classification of the event.

This event-level classification can also serve as an auxiliary task for signal-only datasets. For the $t\bar{t}$ dataset, we found that classifying the number of reconstructable tops (0, 1, or 2) improved performance. However, a similar auxiliary task for the $t\bar{t}H$ dataset did not yield improvements, indicating that the utility of such a task should be evaluated on a case-by-case basis.

The total loss is the weighted sum of the (binary) cross-entropy losses from the two stages of the hierarchical graph-learning $\mathcal{L}_\text{stage\,\{1,2\}}$, the auxiliary task following the encoding step $\mathcal{L}_\text{aux}$, and the event-level classification $\mathcal{L}_\text{event}$ with weights $\lambda_{\{1,2\}}$.
\begin{equation}
    \mathcal{L} = (1-0.5\alpha)\mathcal{L}_\text{stage\,1} + 0.5\alpha\mathcal{L}_\text{stage\,2} +\lambda_1\mathcal{L}_\text{aux}+\lambda_2\mathcal{L}_\text{event}
\end{equation}
The weights were optimised for each training to ensure all components contribute equally. For the $t\bar{t}$ dataset we choose $\lambda_1 = 1/7$ and $\lambda_2 = 1/6$. For the pure $t\bar{t}H$ trainings we chose $\lambda_1 = \lambda_2 = 1/50$ and $\lambda_2 = 1$ for the combined dataset. The weight $\alpha$ is an epoch-dependent function ranging from 0 to 1: at the beginning of training, the network emphasises the first stage of graph learning for stability, and the contribution of the two stages is gradually balanced such that $\alpha =1$ is reached at epoch 40.
The network has a total of 1.1 million parameters, and all training was performed on a single NVIDIA RTX A6000 GPU. Each network was trained for 500 epochs, taking around 30 hours. 

\subsection{Interpretation of network outputs}
\label{sec:algorithm}
To translate the network's probabilistic outputs into a definite set of reconstructed particles, we employ a post-processing algorithm. This algorithm interprets the edge probabilities from both hierarchical stages to build physical objects. The primary constraint of the algorithm is the prevention of double-counting, ensuring each input particle is assigned to at most one final-state object. While the algorithm can be defined for any collection of processes, we illustrate it here using fully hadronic $t\bar{t}$ events for clarity, followed by its extension to the more complex semileptonic $t\bar{t}H$ topology.

\textbf{Fully hadronic $t\bar{t}$ events} For the $t\bar{t}$ topology, the reconstruction is determined by two probability thresholds $p_W$ and $p_t$, for $W$-boson and top quark identification, respectively. The algorithm is a greedy, iterative process: 

\begin{enumerate}[label=\arabic*.]
    \item All potential $W$ candidates (jet pairs) from the first hierarchical stage are sorted in descending order of their predicted probability.
    \item The algorithm iterates through this sorted list. For each candidate, it checks if its probability exceeds $p_W$ and if its constituent jets have not already been assigned to another object.
    \item If the $W$ candidate is valid, the algorithm consults the second-stage predictions to see if this $W$ meta-node forms a top quark with a third, unassigned jet, with a probability exceeding $p_t$.
    \item If a top quark is successfully identified, all three constituent jets are marked as "used". If the $W$ candidate was valid but did not form a top quark, only its two jets are marked as "used".
    \item This process repeats until no remaining $W$ candidates in the list satisfy the $p_W$ threshold.
\end{enumerate}

This procedure inherently prevents double-counting. The thresholds can be optimised to tune efficiency and purity. It is required that $p_W$ is not set lower than the meta-node selection threshold $p_\text{th}$ used during training, as lower-probability W candidates would not have a corresponding meta-node in the second stage. The algorithm does not enforce a fixed number of reconstructed objects, preserving the model's generality.

\textbf{Semi-leptonic $t\bar{t}H$ events} The extension to $t\bar{t}H$ events introduces the Higgs boson and requires an additional probability threshold, $p_H$. We empirically set a stricter cut for the Higgs boson ($p_H > p_W$). Since the first-stage classification is now multi-class (distinguishing between $W$, $H$, etc.), the algorithm is adapted as follows. 

The algorithm considers a single list of all potential $W$ and $H$ candidates, sorted by their maximum class probability. During the iteration, a candidate is accepted only if its probability exceeds the corresponding threshold and its constituents are available. If a candidate is identified, its constituents are marked as used, and it proceeds to the second stage for top quark reconstruction as before. The algorithm treats both hadronic and leptonic $W$ boson reconstructions uniformly. The leptonic $W$, formed by the lepton and MET, is a trivial combination for the network to identify in this event topology.

The generalization to multiprocesses is straightforward by defining the appropriate thresholds for each particle type without modifying the architecture.

\section{Results}
We evaluate the performance of TIGER by comparing its reconstruction efficiency and purity against established baseline models. A direct comparison presents a challenge due to a fundamental difference in design philosophy. Baseline models like HyPER and SPANet are specialised for a known event topology and typically predict a fixed number of objects. In contrast, TIGER is intentionally topology-agnostic, making no a priori assumptions about the event structure or the number of reconstructable particles.
This agnostic approach is designed to be more applicable to realistic experimental conditions, where the event topology is not known in advance. While specialised models may achieve high efficiency by constraining the problem space, this can come at the cost of purity when applied to datasets containing a mix of fully and partially reconstructable events. Our objective is to demonstrate that TIGER's generalised framework achieves efficiencies comparable to specialised models while yielding significantly higher purity.

\subsection{Fully hadronic $t\bar{t}$ events}
For this analysis, we set the probability thresholds to $p_W = p_T = 0.3$. These values were chosen to match the overall top quark efficiency of the HyPER baseline, allowing for a direct comparison of purity. Efficiency for a given particle type is defined as the ratio of correctly reconstructed particles to the total number of reconstructable particles at the truth level ($N^p_\text{correct} / N^p_\text{reconstructable}$). Purity is the ratio of correctly reconstructed particles to the total number of predicted particles ($N^p_\text{correct} / N^p_\text{all pred}$). 
\begin{table}[t]
\centering
\small
\begin{tabular}{r|
    *{3}{>{\centering\arraybackslash}p{.6cm}}|
    *{3}{>{\centering\arraybackslash}p{.6cm}}|
    *{3}{>{\centering\arraybackslash}p{.6cm}}|
    *{3}{>{\centering\arraybackslash}p{.6cm}}}
    \toprule
    & \multicolumn{3}{c|}{\textrm{HyPER} Eff. [\%]} & \multicolumn{3}{c|}{\textrm{TIGER} Eff. [\%]}  & \multicolumn{3}{c|}{\textrm{HyPER} Pur. [\%]} & \multicolumn{3}{c}{\textrm{TIGER} Pur. [\%]}\\[2pt]
    & $t\bar{t}$ & $t$ & $W$ & $t\bar{t}$ & $t$ & $W$ & $t\bar{t}$ & $t$ & $W$ & $t\bar{t}$ & $t$ & $W$ \\
\midrule
 6 j        & 80 & 71 & 75 & 78 & 70 & 75 & 16 & 40 & 49 & \textbf{34} & \textbf{54}  & \textbf{55}\\
 7 j        & 66 & 68 & 71 & 65 & 67 & 72 & 19 & 41 & 49 & \textbf{28} & \textbf{49}  & \textbf{51}\\
 $\geq$8 j  & 51 & 61 & 65 & 50 & 61 & 67 & 18 & 40 & \textbf{47} & \textbf{23} & \textbf{44}  & 42\\
 All        & 66 & \underline{67} & \underline{71} & 65 & \underline{66} & \underline{72} & 18 & 40 & 49 & \textbf{29} & \textbf{49}  & \textbf{50}\\
\bottomrule
\end{tabular}
\caption{Performance metrics for different jet multiplicities, comparing the HyPER and TIGER methods. The efficiencies of HyPER are taken from~\cite{hyper}, and purities are calculated from the dataset using the efficiencies. Underlined entries indicate efficiencies at which TIGER was aligned with HyPER for a fair purity comparison.  Higher purity performance is indicated in bold.}
\label{tab:ttbar}
\end{table}

As summarised in Table \ref{tab:ttbar}, TIGER achieves efficiencies comparable to the baseline while demonstrating significantly higher purities. For single top quarks, our model's purity is nearly 10\% higher on average. The most striking difference is observed in the $t\bar{t}$ event-level purity, where our performance is more than double that of the baseline at low jet multiplicity. This discrepancy arises because only approximately 27\% of events in the dataset contain two fully reconstructable top quarks. Specialised models designed to always predict two top quarks are therefore inherently prone to incorrect reconstructions in the remaining 73\% of events. 

For single $W$ bosons, TIGER shows a 6\% purity improvement for events with 6 jets. At high multiplicity, HyPER achieves better purity, as it inherently limits predictions to the two highest-probability candidates, which artificially reduces the rate of fake reconstructions. While the authors of HyPER note that probability cuts could be implemented, the impact of such a change on their reported performance is not shown.

\subsection{Semi-leptonic $t\bar{t}H$ events}
\begin{figure}[h!]
    \centering
    \includegraphics[width=0.9\linewidth]{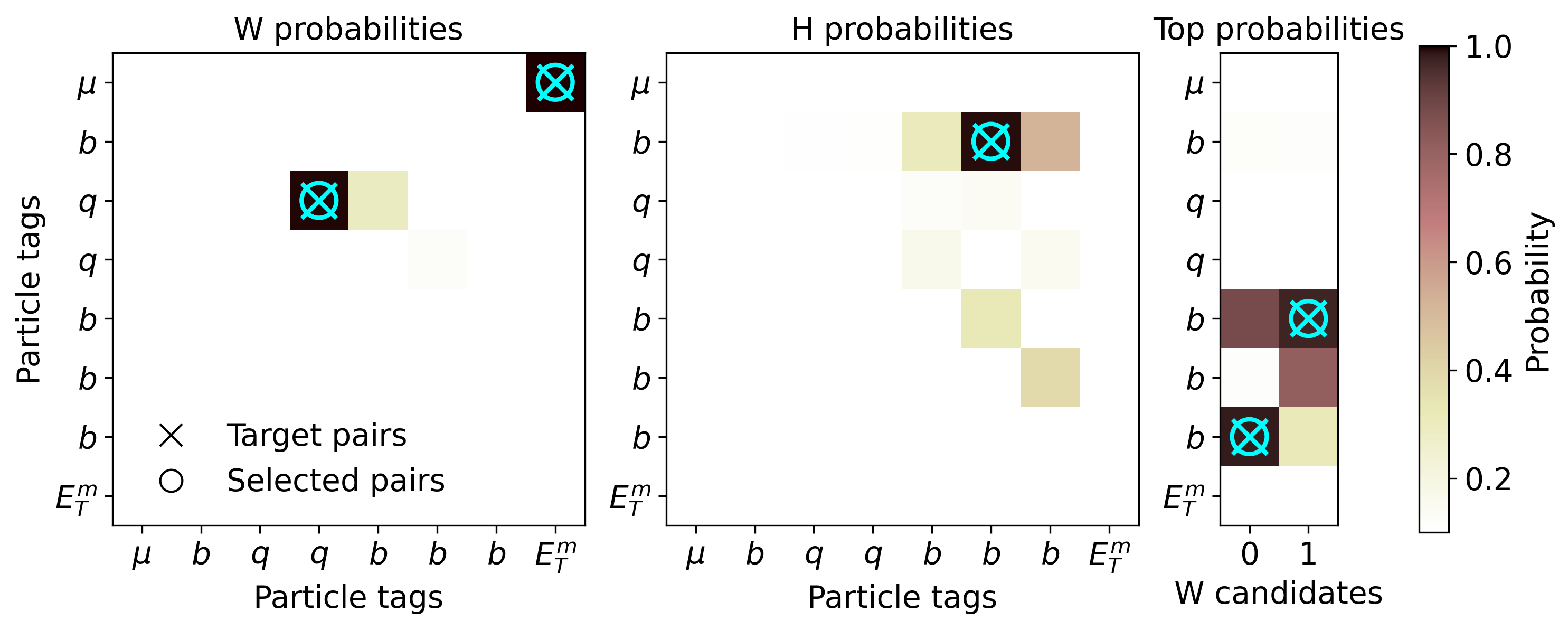}
    
    
    \includegraphics[width=0.9\linewidth]{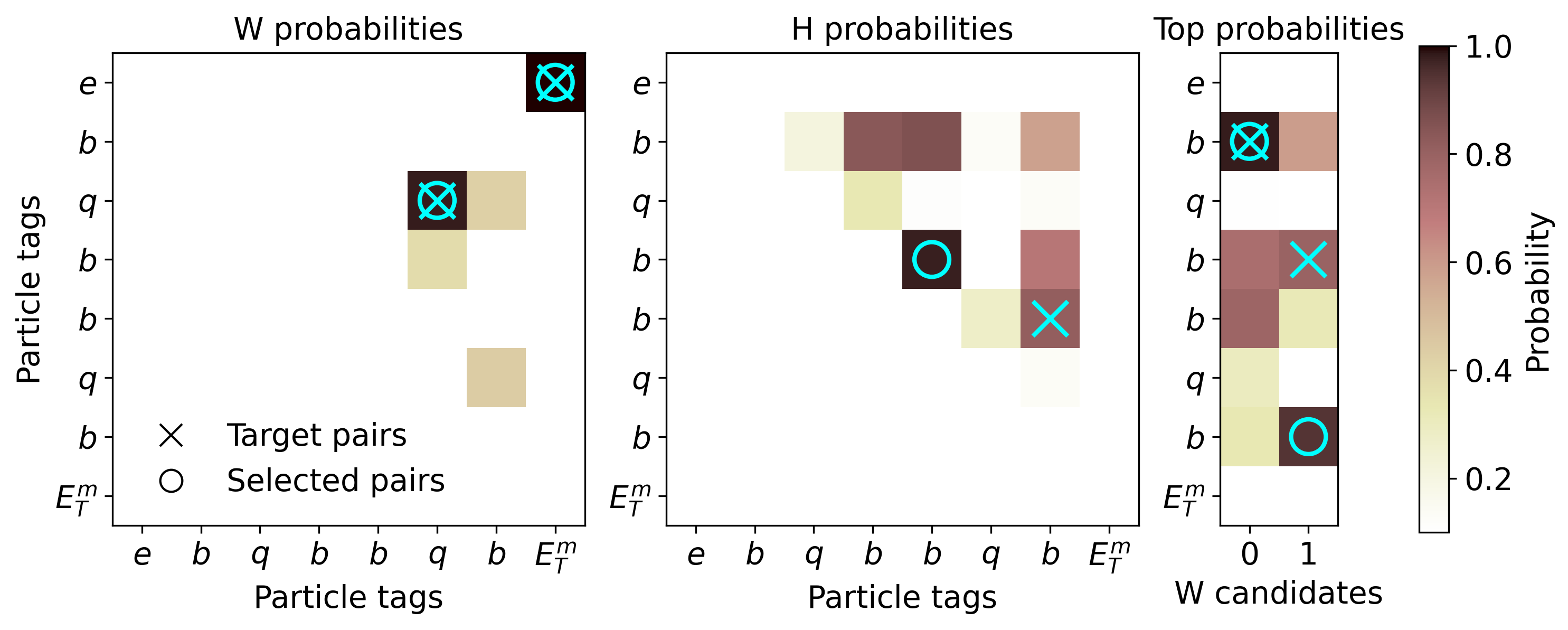}
    
    \caption{Two event displays from the $t\bar{t}H$ test dataset. The left and middle matrices show the classification probabilities from the first stage of the hierarchical graph, while the right matrix depicts the second-stage incidence matrix, showing only the $W$ candidates selected by the first stage. The selected pairs are determined by the algorithm described in Section~\ref{sec:algorithm}. The top panel shows a correctly reconstructed event, whereas the bottom panel illustrates a misassignment between a $b$-jet from the Higgs and the top.}
    \label{fig:events}
\end{figure}

To test our model on a more complex event topology, we use the semileptonic $t\bar{t}H$ dataset and compare our results against SPANet. We set the probability thresholds to $p_W = p_T = 0.3$ and $p_H = 0.9$, with the value of $p_H$ chosen to align our Higgs boson efficiency with that of SPANet for a fair comparison. We note that the Higgs and hadronic top efficiencies are highly sensitive to the $p_H$ threshold, as the primary source of misreconstruction is the confusion between $b$-jets originating from the Higgs and those from the hadronic top decay. Two examplary reconstructed events are displayed in Figure~\ref{fig:events}. While the top-panel event is correctly reconstructed, the bottom panel illustrates the misassignment described above.

\begin{table}[b!]
  \centering
  \small
  \begin{tabular}{r|
    *{4}{>{\centering\arraybackslash}p{.45cm}}|
    *{4}{>{\centering\arraybackslash}p{.45cm}}|
    *{4}{>{\centering\arraybackslash}p{.45cm}}|
    *{4}{>{\centering\arraybackslash}p{.45cm}}}
    \toprule
    & \multicolumn{4}{c|}{\textrm{SPANet} Eff. [\%]} & \multicolumn{4}{c|}{\textrm{TIGER} Eff. [\%]}  & \multicolumn{4}{c|}{\textrm{SPANet} Pur. [\%]} & \multicolumn{4}{c}{\textrm{TIGER} Pur. [\%]}\\[2pt]
    & $t\bar{t}H$ & $t_h$ & $t_l$ & $H$ & $t\bar{t}H$ & $t_h$ & $t_l$ & $H$ & $t\bar{t}H$ & $t_h$ & $t_l$ & $H$ & $t\bar{t}H$ & $t_h$ & $t_l$ & $H$  \\
    \midrule
    6 j        & 54 & 54  & 69  & 50 & 50  & 51  & 70  & 51 & 15  & 24  & 63  & 38 & \textbf{32}  & \textbf{35}  & \textbf{64}  & \textbf{50}   \\
    7 j        & 42 & 48  & 68  & 49 & 39  & 46  & 69  & 48 & 16  & 26  & 63  & 38 & \textbf{27}  & \textbf{34}  & \textbf{64}  & \textbf{47}   \\
    $\geq$8 j  & 33 & 42  & 68  & 47 & 30  & 40  & 67  & 45 & 14  & 26  & \textbf{63}  & 37 & \textbf{22}  & \textbf{31}  & 62  & \textbf{42}   \\
    All        & 45 & 49  & 69  & \underline{49} & 40  & 46  & 69  & \underline{49} & 15  & 25  & 63  & 38 & \textbf{27}  & \textbf{34}  & \textbf{64}  & \textbf{47}   \\
    \bottomrule
  \end{tabular}
  \caption{Efficiency and purity comparison for $t\bar{t}H$ events. TIGER and the SPANet baseline are evaluated across varying jet multiplicities. Reconstruction targets include the full event ($t\bar{t}H$), hadronic top ($t_h$), leptonic top ($t_l$), and Higgs boson ($H$). SPANet values were taken from~\cite{spanet2}. Underlined entries indicate efficiencies at which TIGER was aligned with SPANet for a fair purity comparison.  Higher purity performance is indicated in bold.}  
  \label{tab:tth}
\end{table}

The performance metrics are presented in Table~\ref{tab:tth}. While efficiencies for the semileptonic top and Higgs boson are comparable, our model slightly underperforms in reconstructing the hadronic top and the full event. However, TIGER again demonstrates a substantial improvement in purity, outperforming the baseline by approximately 10\% on average for hadronic tops, Higgs bosons, and full-event reconstructions.

This performance is achieved despite dataset imbalances between the training and testing sets due to the different $b$-tagging requirements. These differences likely pose a greater challenge for our agnostic approach, which does not presuppose a specific event structure. While later work has explored applying probability cuts to SPANet outputs~\cite{spanet3}, its architecture remains fundamentally constrained by the predefined event topology. For example, it could never predict two Higgs bosons in an event, which is explicitly allowed in our model, further enhancing its generality.

\subsection{Classification of $t\bar{t}H$ and $t\bar{t}+bb$}
\begin{figure}[t]
    \centering
    \includegraphics[width=0.6\textwidth]{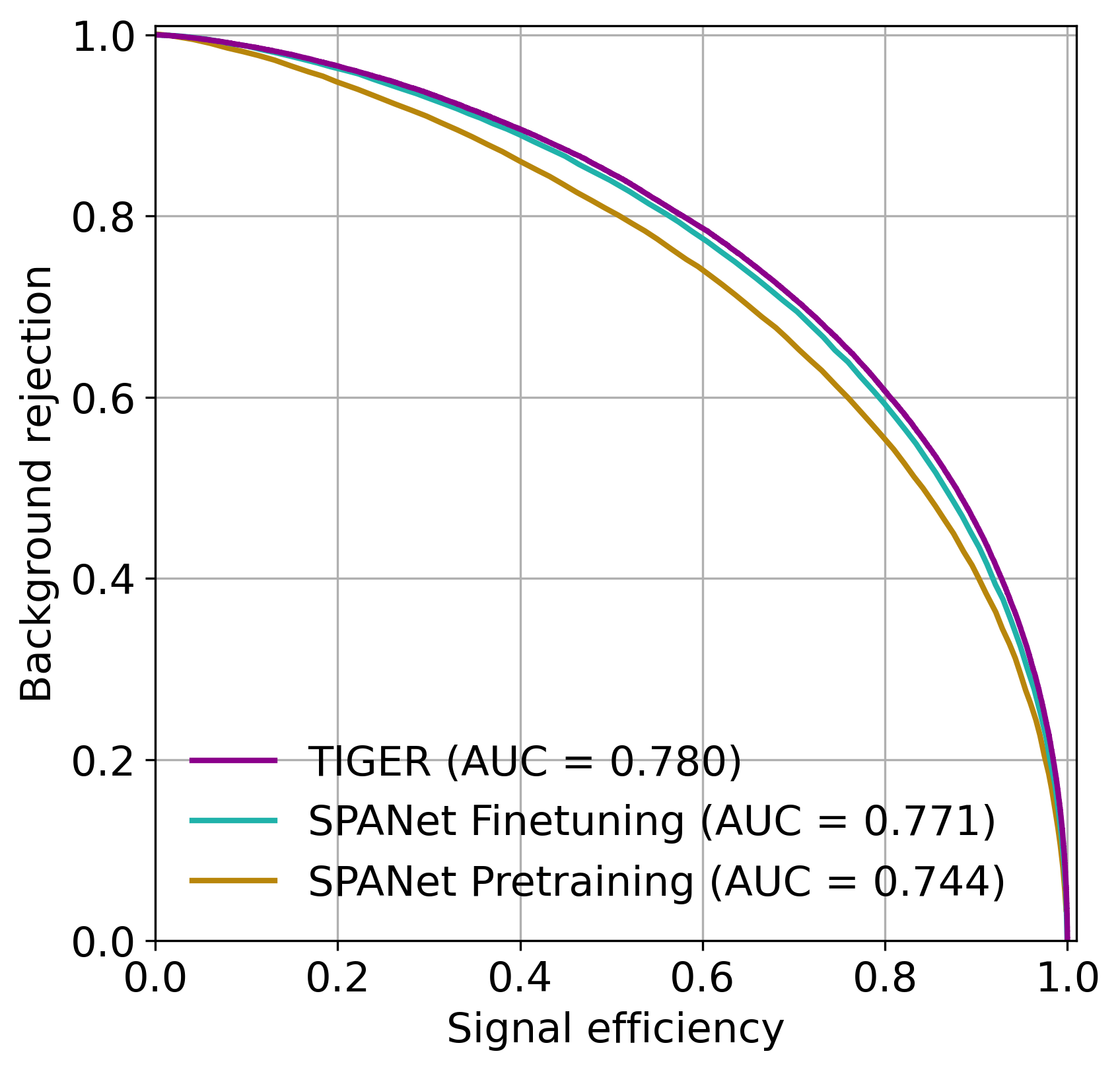}
    \caption{Receiver Operating Characteristic (ROC) curves for the $t\bar{t}H$ vs. $t\bar{t}+bb$ classification, comparing TIGER with SPANet baselines (results taken from~\cite{spanet2}).}
    \label{fig:roc}
\end{figure}

Finally, we evaluate the model's capability for signal-versus-background discrimination. Using the combined $t\bar{t}H$ and $t\bar{t}+bb$, we train TIGER with the event classification head simultaneously with the reconstruction components. This end-to-end approach is compared to SPANet, which also uses a global event representation for classification. The authors of~\cite{spanet2} found that fine-tuning a pre-trained SPANet backbone for classification yielded the best performance. As shown by the ROC curves in Figure~\ref{fig:roc}, TIGER outperforms even the optimised fine-tuning strategy, demonstrating its effectiveness as a multi-task learning framework.
\section{Conclusions}
In this work, we introduced TIGER, a novel, topology-agnostic method for event reconstruction. Its hierarchical graph architecture is inspired by the fundamental physics principle of sequential two-body particle decays. Unlike specialised models such as HyPER and SPANet, which require a predefined event topology as an architectural prior or direct input, TIGER makes no such assumptions during either training or evaluation. This makes the learning task more challenging but yields a far more flexible and realistic framework for real-world physics analysis.

On top of being topology agnostic, we demonstrated on fully hadronic $t\bar{t}$ and semileptonic $t\bar{t}H$ events that this approach also achieves better performance compared to specialized baselines. This is reflected in the fact that, for similar reconstruction efficiencies, we deliver significantly higher purity. The improvement in purity is a direct consequence of the model’s agnosticism; it avoids the inherent bias of methods that are optimised to find a complete event structure, even when many events are not fully reconstructable. Furthermore, we showed that TIGER's architecture seamlessly integrates event classification, outperforming a fine-tuned baseline in a signal-versus-background task without requiring a multi-stage training process. Collectively, these results establish TIGER as a powerful and flexible tool for data-driven event reconstruction.

The performance demonstrated in this study is primarily limited by the provided datasets. Future work will focus on deploying the model with more realistic detector simulations to better assess its performance under experimental conditions. The model's flexibility allows for the natural inclusion of additional information, such as jet substructure variables or outputs from advanced flavour-tagging algorithms, which could further boost performance.

Ultimately, this approach could be extended to complex analyses involving multiple overlapping signal and background topologies, such as searches for rare $H\rightarrow c\bar{c}$ decays~\cite{cms_tthcc,vhbbcc_atlas} or Di-Higgs events~\cite{dihiggs_comb}, paving the way toward a unified model capable of addressing diverse physics analyses at the LHC.
\section*{Code availability}
The code is publically available at \url{https://github.com/nathaliesoy/tiger}.
\section*{Acknowledgments}
We thank Alberto Ibanez, Zihan Zhang, Alexander Shmakov, and Hideki Okawa for insightful discussions and clarifications concerning the baseline models used in our comparisons.
NS, NK, and EG are supported by the Minerva Stiftung under grant number 715027, and the Weizmann Institute for Artificial Intelligence grant program Ref 151676. 
\printbibliography
\end{document}